\documentstyle[aps,times,epsf,12pt]{revtex}
\newcommand{\tfrac}[2]{{\textstyle\frac{#1}{#2}}}

\begin{document} 
\title{Quantum Statistical Mechanics of Nonrelativistic Membranes: \\
  Crumpling Transition at Finite Temperature}
%
\tighten
\author{M. E. S. Borelli\thanks{E-mail: borelli@physik.fu-berlin.de}, H.
  Kleinert\thanks{E-mail: kleinert@physik.fu-berlin.de; web site:
    http://www.physik.fu-berlin/$\sim$kleinert}, and Adriaan M. J.
  Schakel\thanks{E-mail: schakel@physik.fu-berlin.de}} \address{Institut f\"ur
  Theoretische Physik \\ Freie Universit\"at Berlin \\ Arnimallee 14, 14195
  Berlin.}  \date{\today} \maketitle

\begin{abstract}
  The effect of quantum fluctuations on a nearly flat, nonrelativistic
  two-dimensional membrane with extrinsic curvature stiffness and
  tension is investigated.  The renormalization group ana\-lysis is
  carried out in first-order perturbative theory.  In contrast to
  thermal fluctuations, which soften the membrane at large scales and
  turn it into a crumpled surface, quantum fluctuations are found to
  {\em stiffen} the membrane, so that it exhibits a Hausdorff
  dimension equal to two.  The large-scale behavior of the membrane is
  further studied at finite temperature, where a nontrivial fixed
  point is found, signaling a crumpling transition.
\end{abstract}


\section{Introduction} \label{intro}

Quantum oscillations, analogous to the superconducting Josephson effect,
have recently been detected in samples of superfluid $^3$He \cite{qm1,qm2}.
The apparatus involved in these experiments consists of an inner cell,
filled with $^3$He, contained in an outer cell, also filled with $^3$He.
The two containers are separated by a stiff membrane glued to the bottom of
the inner cell, and by a softer one attached to its top.  The lower membrane
contains an array of small apertures allowing for exchange of atoms between
the cells, equivalent to the superconducting weak link.  By manipulating
this membrane, the pressure between the two systems can be kept at a fixed
value, and the resulting mass current is determined by the displacement of
the upper membrane from its original position.  Josephson current
oscillations at the weak link are then observed as oscillations of this
membrane.

It is thus of interest to investigate whether the membranes themselves
display quantum fluctuations, and to what extent such effects may be
observable.

Thermal fluctuations are known to soften fluid membranes, and to
increase their surface tension
\cite{helf3,peliti,klein1,forster,helf2,forster2,fluid,cai}.  At a
temperature $T$, the membrane's bare surface tension $r_0$ and ben\-ding
rigidity $1/\alpha_0$ are renormalized at large scales as follows:
\begin{eqnarray} 
\label{rold}
r &=& r_0 \left[ 1 + \frac{1}{4 \pi} u_{{\rm th},0} \ln (\Lambda L )
\right], \\
\label{kappa}
\frac{1}{\alpha} &=& \frac{1}{\alpha_0} \left[ 1 - \frac{3}{4 \pi}
u_{{\rm th},0} \ln (\Lambda L) \right] ,
\end{eqnarray} 
where 
\begin{equation} \label{uth}
u_{{\rm th},0} = k_{\rm B} T \alpha_0
\end{equation}  
is the (dimensionless) expansion parameter in the regime dominated by
thermal fluctuations, while $\Lambda$ is an ultraviolet wavenumber
cutoff set by the inverse size of the molecules in the membrane, and $L$
is an infrared cutoff determined by its finite size.  The
renormalization group equation extracted from Eq.\ (\ref{kappa}) yields
only the trivial fixed point $\alpha=0$, implying the absence of a
crumpling transition.  The fixed point turns out to be unstable in the
infrared.  As a result, the renormalized coupling constant $\alpha$
flows away from it, so that the bending rigidity $1/\alpha$ tends to
zero with increasing membrane size.  The length scale where the
right-hand side of Eq.\ (\ref{kappa}) vanishes gives the so-called de
Gennes-Taupin persistence length $\xi_{{\rm th},0} = \Lambda^{-1} \exp
(4 \pi / 3 u_{{\rm th},0})$ \cite{DGT}, above which the bending rigidity
$1/\alpha$ vanishes.  As a consequence, such membranes are crumpled at
these scales, and fill the space in which they are embedded completely.
Such objects are said to have infinite Hausdorff dimension $d_{\rm H} =
\infty$.  The full SO($d$) rotational symmetry of the $d$-dimensional
embedding space is recovered in this way.  For a smooth surface, this
symmetry would be spontaneously broken to its SO($d-2$) $\times$ SO(2)
subgroup.  The absence of this symmetry breaking is in accord with the
Mermin-Wagner theorem \cite{MW}, stating that in a two-dimensional
system long-range order is destroyed at any finite temperature.

The results (\ref{rold}) and (\ref{kappa}) are derived from the
Canham-Helfrich model \cite{canham,helf1}, describing stiff membranes
subject to thermal undulations.  The Hamiltonian reads:
\begin{equation}
{\cal H}_0 = \int {\rm d}S \left(r_0 \ + \frac{1}{2\alpha_0} H^2 \right),
\label{helfmodel}
\end{equation} 
where ${\rm d}S$ is the surface element, and $H$ corresponds to (twice) the
mean curvature of the surface at each point.  We omit a term proportional to
the Gaussian curvature since it is only important if fluctuations change the
topology of the membrane, characterized by the Euler characteristic, i.e.,
by the number of handles---a possibility which we shall neglect.  Each point
on the membrane is represented by a vector ${\bf X}(\sigma_1,\sigma_2)$ in
the $d$-dimensional bulk space, depending on two coordinates $\sigma_1$ and
$\sigma_2$.  Explicitly, the Hamiltonian (\ref{helfmodel}) reads (for
reviews, see Ref.\ \cite{Jerusalem})
\begin{equation}
{\cal H}_0 =\int {\rm d}^2 \sigma \sqrt{g} \left[r_0 + \frac{1}{2 \alpha_0}
(\Delta {\bf X})^2 \right]
\label{helfmodel2},
\end{equation} 
where 
\begin{equation} \label{gmunu}
g_{a b}= \partial_{a}{\bf X} \cdot \partial_{b}{\bf X}
\end{equation} 
is the metric induced by the embedding and $g \equiv \det[g_{a b}]$.
The symbol $\partial_a \, (a=1,2)$ denotes the derivative with respect
to the coordinates $\sigma = (\sigma_1, \sigma_2)$, and $\Delta = g^{a
b}D_a D_b$ is the scalar Laplacian, where $D_a$ is the covariant
derivative associated with the metric.

By lowering the temperature, the effect of thermal fluctuations decreases,
and the behavior of the membrane becomes dominated by quantum effects.  To
account for these, we have to include time-dependence and add a kinetic term
to the classical theory.  For an incompressible surface, the kinetic term
reads, in euclidean spacetime,
\begin{equation}
{\cal T} = \frac{1}{2 \nu_0} \int {\rm d}^2 \sigma \sqrt{g} \, \dot{\bf
X}^2,
\label{dynterm}
\end{equation}  
where $\bf X$ is now time-dependent, $1/ \nu_0$ is the bare mass density, and
the dot indicates a time derivative.

The partition function $Z$ can be represented as a functional integral over
all possible surface configurations ${\bf X}(\sigma, \tau)$:
\begin{equation} \label{partfunc2}
Z = \int {\cal D}{\bf X} \exp(-S_0[{\bf X}]/\hbar),
\end{equation}
with the euclidean action 
\begin{equation} \label{action}
S_0  =  \int {\rm d} \tau \, {\rm d}^2 \sigma \sqrt{g} \left[
\frac{1}{2\nu_0} \dot{\bf X}^2 + r_0 + \frac{1}{2 \alpha_0}
(\Delta {\bf X})^2 \right].
\label{euclaction}
\end{equation} 
We shall in this paper study the quantum statistical mechanics of a
membrane described by this action.  

Specifically, in Sec.\ {\ref{1loop}}, we give a one-loop analysis of the
effect of quantum fluctuations on such a two-dimensional surface
embedded in $d=3$ space dimensions.  We show that, contrary to thermal
fluctuations, quantum fluctuations lead to a stif\-fening of the
membrane at large scales.  In the one-loop approximation, the flow
equations are again found to yield only the trivial fixed point $\nu = r
= \alpha =0$, implying the absence of a crumpling transition.  However,
this fixed point is now, in contrast to the one generated by thermal
fluctuations, stable in the infrared.  This means that $\alpha$ flows
towards this fixed point and the bending rigidity tends to infinity with
increasing membrane size, so that the surface remains two-dimensional.

We further investigate the behavior of the system at finite
temperature.  We find at the one-loop order a nontrivial fixed point
arising here.  Being unstable in the infrared, it gives rise to a
crumpling transition.  Below a critical temperature $T_{\rm c}$,
$\alpha$ flows to zero, implying a flat membrane, while above $T_{\rm
c}$, the flow is towards infinity, implying a crumpled membrane.

\section{Perturbative Expansion} \label{1loop}

We shall work in the Monge parametrization, where a point on the surface
embedded in three-dimensional space is described by a displacement field
$\phi({\bf \sigma},\tau)$ with respect to a reference plane ${\bf
\sigma}=(\sigma_1,\sigma_2)$, such that
\begin{equation} 
{\bf X}({\bf \sigma},\tau) =
\left({\bf \sigma},\phi({\bf \sigma},\tau)\right) .
\label{monge}
\end{equation}
Inserting this parametrization in the action (\ref{action}), and
expanding the resulting expression up to fourth order in powers of the
displacement field $\phi$, we obtain
\begin{eqnarray} \label{pertmodel}
S_0 &=& \int {\rm d} \tau {\rm d}^2\sigma \left\{
\frac{1}{2}\left[ \frac{1}{\nu_0} {\dot \phi}^2 + r_0 (\partial_a
\phi)^2 + \frac{1}{\alpha_0} (\partial^2\phi)^2 \right] + \frac{1}{4
\nu_0} {\dot \phi}^2 (\partial_a \phi)^2 - \frac{r_0}{8} (\partial_a
\phi )^2( \partial_b \phi)^2 \right. \\&& \nonumber \hspace{1.8cm}
\left.  - \frac{1}{4 \alpha_0} (\partial_a \phi)^2(\partial^2\phi)^2 -
\frac{1}{\alpha_0} (\partial_a \phi )(\partial_b \phi)( \partial_a
\partial_b \phi)(\partial^2\phi) \right \}.
\end{eqnarray}
The displacement field describes the undulations of the surface.  Its
spectrum, $\omega^2 = c_{\rm s}^2 q^2$, is gapless with $c_{\rm s}^2 = r_0
\nu_0$ the velocity of the transversal waves.

In the one-loop approximation, the exponent in (\ref{partfunc2}) may
be expanded up to second order around a background configuration
$\Phi({\bf \sigma},\tau)$ extremizing $S_0$.  The resulting integral
is Gaussian and yields the effective action
\begin{equation} \label{effaction}
S_{\rm eff}[\Phi] = S_0 [\Phi] + S_1 [\Phi] = S_0 [\Phi] +
\frac{\hbar}{2} \mbox{Tr} \ln \left[\left. \frac{\delta^2 S_0}{\delta
\phi({\bf \sigma},\tau) \delta \phi({\bf \sigma'},\tau')} \right|_{\Phi}
\right],
\end{equation} 
where the expression in square brackets is a functional matrix given by
the second functional derivative of $S_0$, and Tr denotes the functional
trace, i.e., the integral $\int {\rm d} \tau \, {\rm d}^2 \sigma$, as
well as the integral $\int {\rm d} \omega \, {\rm d}^2 q/(2 \pi)^3$ over
the (angular) frequency $\omega$ and the wavevector ${\bf q}$.

Using a derivative-expansion method due to Fraser \cite{caroline}, we
expand the one-loop correction $S_1$ in Eq.\ (\ref{effaction}) in powers of
the derivatives of the field $\phi({\bf \sigma},\tau)$, as in Ref.\
\cite{memb}.  To obtain the renormalization of the parameters $r_0$,
$\alpha_0$ and $\nu_0$, it suffices to keep only the first three terms of
the expansion:
\begin{equation}\label{seff}
S_1 = \frac{\hbar}{2} \int {\rm d}\tau \, {\rm d}^2\sigma \left[ I_1
{\dot \phi}^2 + I_2 (\partial_a \phi)^2 + I_3 (\partial^2\phi)^2 +
\ldots \right],
\end{equation}
with
\begin{eqnarray}
I_1 &=&  \frac{1}{2\nu_0} \int \frac{{\rm d}\omega}{2\pi} 
\frac{{\rm d}^2q}{(2\pi)^2} \frac{q^2}{\omega^2/\nu_0 + r_0 q^2 +
q^4/\alpha_0}
\label{I1} \\
I_2 &=& \int \frac{{\rm d}\omega}{2\pi} \frac{{\rm d}^2q}{(2\pi)^2}
\frac{\frac{1}{2} \omega^2/\nu_0 -r_0 q^2 - \frac{3}{2}
q^4/\alpha_0}{\omega^2/\nu_0 + r_0 q^2 + q^4/\alpha_0}
\label{I2} \\
I_3 &=& -\frac{3}{2} \frac{1}{\alpha_0} \int \frac{{\rm d}\omega}{2\pi} 
\frac{{\rm d}^2q}{(2\pi)^2} \frac{q^2}{\omega^2/\nu_0 + r_0 q^2 + 
q^4/\alpha_0}.
\label{I3}
\end{eqnarray}
After the integrals over the loop energy $\omega$ have been carried out, the
resulting momentum integrals in Eqs.\ (\ref{I1})--(\ref{I3}) diverge in the
ultraviolet.  To regularize them, we introduce a wavenumber cutoff
$\Lambda$.  Contributions proportional to positive powers of $\Lambda$ are
irrelevant, and will be ignored.  In dimensional regularization, where the
number $D$ of space dimensions of the membrane is ana\-lytically continued
to be less than two, $D=2-\epsilon$, these powerlike divergences never
appear in the first place.  Only logarithmic divergences arise as poles in
$1/\epsilon$.  Substituting the results of the integration into Eq.\
(\ref{seff}), we obtain the effective action
\begin{equation} \label{seff2}
S_{\rm eff} = \frac{1}{2} \int {\rm d}\tau \, {\rm d}^2 \sigma \left[
\frac{1}{\nu} {\dot \phi}^2 + r (\partial_a \phi)^2 + \frac{1}{\alpha}
(\partial^2\phi)^2 + \ldots \right],
\end{equation}
with the renormalized inverse mass density $\nu$, surface tension $r$, and
inverse rigidity $\alpha$
\begin{eqnarray} 
\frac{1}{\nu} &=& \frac{1}{\nu_0} \left[ 1 -
\frac{1}{16 \pi}u_{{\rm qm},0} \ln\left(\Lambda \xi_{{\rm qm},0}
\right) \right], \label{zetaeff} \\ r &=& r_0 \\
\frac{1}{\alpha} &=& \frac{1}{\alpha_0} \left[ 1 +
\frac{3}{16\pi}u_{{\rm qm},0} \ln\left(\Lambda \xi_{{\rm qm},0}
\right) \right], \label{kappaeff}
\end{eqnarray}
where the dimensionless parameter $u_{{\rm qm},0} = \hbar r_0
\alpha_0^{3/2} \nu_0^{1/2}$ is the (bare) expansion parameter in the
quantum regime.  Note that the surface tension is not renormalized by
quantum fluctuations at this order.  The parameter $\xi_{{\rm qm},0} =
2/\sqrt{r_0 \alpha_0}$ in the argument of the logarithm in Eqs.\
(\ref{zetaeff}) and (\ref{kappaeff}) defines a characteristic length
scale of the problem.  It sets the scale at which the tension and
stiffness terms in the action (\ref{seff2}) become equally important.
At larger scales, the second term in the expansion (\ref{seff2}) becomes
more important and the undulations are dominated by tension, while at
smaller scales, the third term dominates and the undulations are
controlled by stiffness.

To obtain the flow equations, we apply Wilson's procedure
\cite{Wilson}.  Integrating out a momentum shell $\Lambda/s < q <
\Lambda$, rescaling the coupling constants $\hat{\nu} = s^{\epsilon-3}
\nu, \;\; \hat{r} \to s^{-(\epsilon-3)} r, \;\; \hat{\alpha} \to
s^{\epsilon-1} \alpha$, we arrive at
\begin{eqnarray} \label{bnu}
\beta_\nu({\nu},{r},{\alpha}) &=& s \frac{\partial
\hat{\nu}}{\partial s} \Biggl|_{s=1} = (\epsilon-3) {\nu} +
\frac{1}{16\pi} {u}_{\rm qm} {\nu}, \\
\beta_r({\nu},{r},{\alpha}) &=& s \frac{\partial
\hat{r}}{\partial s} \Biggl|_{s=1} = -(\epsilon-3) {r}, \\
\beta_\alpha({\nu},{r},{\alpha}) &=& s
\frac{\partial \hat{\alpha}}{\partial s} \Biggl|_{s=1} = (\epsilon -1)
{\alpha} -\frac{3}{16\pi} {u}_{\rm qm} {\alpha},
\label{betakappa}
\end{eqnarray}
where $\epsilon = 2-D$ is assumed to be small and will be set to zero
at the end of the calculation.  The coefficients of the first terms at
the right-hand sides denote the scaling dimension of the scaling
fields.  For small $\epsilon$, the scaling fields ${\nu}$ and
${\alpha}$ are irrelevant, while ${r}$ is relevant.
Criticality is obtained by setting the relevant fields to zero, i.e.,
${r}=0$ in this case.  Starting somewhere on the critical surface
${r}=0$, the system flows towards the trivial fixed point
${\nu}={\alpha}=0$.  This guarantees the stiffness of the
membrane at large scales.  There is no crumpling transition at the
absolute zero of temperature; the membrane is always flat, where the
normal vectors to its surface are strongly correlated.  More
specifically, the correlation function between the normal vectors to
the surface behaves at large scales as
\begin{equation} \label{algebraic}
\langle \partial_a {\bf X}(\sigma,\tau) \cdot \partial_b {\bf X}(\sigma',\tau)
\rangle \sim \frac{\delta_{a b}}{|\sigma - \sigma'|^3}.
\end{equation}
This algebraic fall-off implies the absence of a persistence length
which would define the length scale above which the normals become
uncorrelated and the surface becomes crumpled.

To investigate this further, let us calculate the Hausdorff dimension
$d_{\rm H}$ of the membrane.  It can be defined by the relation between its
mean surface area $\langle A \rangle$, where
\begin{equation} 
A = \int {\rm d}^2 \sigma \sqrt{g},
\end{equation} 
and the frame, or projected area $A_0 = \int {\rm d}^2 \sigma$.  This
relation is
\begin{equation} 
\langle A \rangle \sim A_0^{d_{\rm H}/2}, \label{defhausd}
\end{equation} 
so that the Hausdorff dimension is given by
\begin{equation} 
d_{\rm H} = 2 \frac{\partial \ln \langle A \rangle}{\partial \ln A_0}.
\label{hausdorff}
\end{equation}
Since the frame area $A_0$ scales with the cutoff $\Lambda$ as
\begin{equation} 
A_0 = \int {\rm d}^2 \sigma \sim \Lambda^{-2},
\end{equation} 
Eq.\ (\ref{hausdorff}) can also be written in the form
\begin{equation} \label{alternative}
d_{\rm H} = - \frac{\partial \ln \langle A
\rangle}{\partial \ln \Lambda}.
\end{equation} 
At the one-loop level, the mean surface area is
\begin{eqnarray} 
\langle A \rangle &=& \left\langle \int {\rm d}^2 \sigma \left[ 1 +
\frac{1}{2} (\partial_a \phi)^2 + \ldots \right] \right\rangle \nonumber
\\ &=& A_0 \left( 1 + \frac{\hbar}{2} \int \frac{{\rm d}\omega}{2\pi}
\frac{{\rm d}^2q}{(2\pi)^2} \frac{ q^2}{ \omega^2/\nu_0 + r_0 q^2 +
q^4/\alpha_0} \right), \label{area}
\end{eqnarray}  
so that we obtain from relation (\ref{alternative}) the Hausdorff
dimension
\begin{equation}
d_{\rm H} = 2 + \frac{1}{16 \pi} {u}_{\rm qm}.
\end{equation}
For large membranes, ${u}_{\rm qm} \to 0$, implying a Hausdorff dimension
$d_{\rm H} =2$.  Expressed in group theoretic terms, the SO($d$)
rotational symmetry of $d$-dimensional space is spontaneously broken to
its SO($d-2$) $\times$ SO(2) subgroup.  Since we are now at zero
temperature, where the membrane has an extra (time) dimension, this
spontaneous symmetry breaking does not violate the Mermin-Wagner
theorem.  The algebraic decay found in Eq.\ (\ref{algebraic}) is an
immediate consequence of this symmetry breaking and identifies the two
resulting Goldstone modes.

We next investigate how the high-temperature regime dominated by thermal
fluctuations, where the membrane is found to be always crumpled, goes over
into the low-temperature regime dominated by quantum fluctuations, where the
membrane remains flat.  To investigate this temperature dependence, we adopt
the imaginary-time approach to thermal field theory \cite{Rivers,Kapusta}.
It can be derived from the corresponding (euclidean) quantum theory at zero
tempe\-rature by restricting the euclidean time to the finite interval $0
\leq \tau \leq \hbar/ k_{\rm B}T$, and substituting
\begin{equation} \label{fun:sub}
\int \frac{{\rm d} \omega}{2\pi}\,g(\omega)\rightarrow \frac{k_{\rm
B}T}{\hbar} \sum_{n} \, g(\omega_{n}),
\end{equation} 
where $g$ is an arbitrary function, and  $\omega_n$ denote the Matsubara
frequencies,
\begin{equation} \label{fun:matb}
\omega_n = 2 \pi n k_{\rm B}T/\hbar , \;\;\;\;\; n = 0, \pm 1, \pm 2 \cdots.
\end{equation} 

Using this substitution in the zero-temperature integrals
(\ref{I1})--(\ref{I3}), as well as the formula \cite{gradstein}
\begin{equation}
\frac{k_{\rm B} T}{\hbar} \sum_n \frac{1}{\omega_n^2 + a^2} = \frac{1}{2
a} \coth\left(\frac{\hbar a}{2 k_{\rm B} T}\right)
\end{equation}
to carry out the sum over the Matsubara frequencies, we arrive at
\begin{eqnarray}
\label{nu}
\frac{1}{\nu}&=& \frac{1}{\nu_0}\left [1  + \frac{1}{8 \pi}
u_{{\rm qm},0} F_1(T,\Lambda') \right], \\ 
\label{r}
r &=& r_0 \left[ 1 - \frac{1}{2 \pi} u_{{\rm qm},0} F_2(T,\Lambda') \right], \\
\label{alpha}
\frac{1}{\alpha} &=& \frac{1}{\alpha_0}\left [1 - \frac{3}{8 \pi}
u_{{\rm qm},0} F_1(T,\Lambda') \right],
\end{eqnarray}
with 
\begin{eqnarray}
F_1(T,\Lambda')&=& \int^{\Lambda'} {\rm d}q' \frac{q'^2 \coth
(\tfrac{1}{2} \gamma_0 q' \sqrt{1 + q'^2} )}{\sqrt{1 + q'^2}}
\label{F2} , \\
F_2(T,\Lambda')&=& \int^{\Lambda'} {\rm d}q' \frac{ (\tfrac{3}{4} q'^2
+ q'^4) \coth(\tfrac{1}{2} \gamma_0 q' \sqrt{1 + q'^2})}{\sqrt{1 +
q'^2}}. \label{F1}
\end{eqnarray}
Here, we rescaled the integration variables: $q'=q \xi_{{\rm
qm},0}/2$, $\Lambda' = \Lambda \xi_{{\rm qm},0}/2$, and
$\gamma_0$ stands for $\gamma_0 = \hbar r_0 \alpha_0^{1/2} \nu_0^{1/2}
/k_{\rm B} T $.  This dimensionless parameter can be expressed as the
ratio of the expansion parameters in the quantum and classical regime,
\begin{equation} 
\gamma_0 = u_{{\rm qm},0}/u_{{\rm th},0}.
\end{equation} 

The integrals in Eqs.\ (\ref{F1}) and (\ref{F2}) diverge in the
ultraviolet as $\Lambda' \to \infty$.  As before, we disregard powerlike
divergences, and consider only the logarithmically diverging terms.  We
thus arrive at:
\begin{eqnarray} 
\label{nu'}
\frac{1}{\nu} &=& \frac{1}{\nu_0}\left [1 + \frac{1}{4 \pi} \left(u_{{\rm
th},0} - \frac{1}{4} u_{{\rm qm},0} \right)\ln (\Lambda') \right], \\
\label{r'} 
r &=& r_0 \left[ 1 + \frac{1}{4 \pi} u_{{\rm th},0} \ln(\Lambda') \right],
\\
\label{alpha'} 
\frac{1}{\alpha} &=& \frac{1}{\alpha_0}\left [1 - \frac{3}{4 \pi}
\left(u_{{\rm th},0} - \frac{1}{4} u_{{\rm qm},0} \right) \ln (\Lambda')
\right].
\end{eqnarray}
One may check that as the temperature $T$ and, consequently, $u_{{\rm th},0}
\propto T$ tend to infinity, Eqs.\ (\ref{r'}) and (\ref{alpha'}) reproduce,
up to finite terms, the high-temperature results (\ref{rold}) and
(\ref{kappa}).  On the other hand, in the limit $u_{{\rm th},0} \propto T
\to 0$, we recover the zero-temperature results
(\ref{zetaeff})--(\ref{kappaeff}).

\begin{figure}
\begin{center}
\epsfxsize=20.cm \mbox{\epsfbox{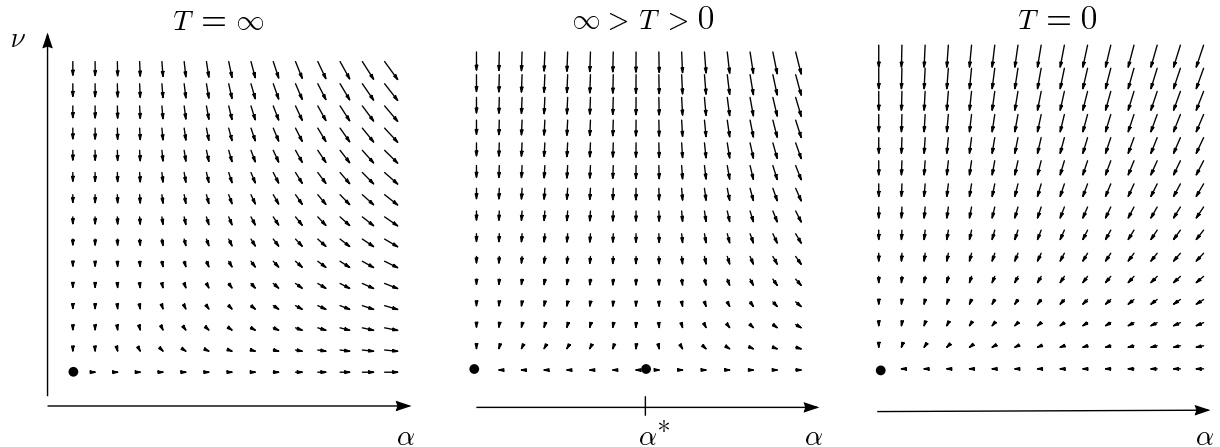}}
\end{center}
\caption{Flow diagrams in the $({\alpha},{\nu})$-plane.  As $T$
becomes finite (middle panel), a nontrivial fixed point (${\alpha}^*
\neq 0$) starts to move to the right away from the origin, and disappears at
infinity when $T$ tends to zero (right panel). \label{fig:flows}}
\end{figure}

In order to explore the behavior of the membrane at large length scales,
we compute the flow equations corresponding to the three parameters of
the theory, as we did above.  They are given by:
\begin{eqnarray}
\label{betanu}
\beta_\nu({\nu},{r},{\alpha}) &=& (\epsilon -3) {\nu}
-\frac{1}{4 \pi}\left({u}_{\rm th} - \frac{1}{4} {u}_{\rm qm} \right)
{\nu}, \\
\label{betar}
\beta_r({\nu},{r},{\alpha}) &=& -(\epsilon -3){r} +
\frac{1}{4 \pi} {u}_{\rm th} {r} , \\
\label{betaalpha}
\beta_\alpha({\nu},{r},{\alpha}) &=& (\epsilon -1)
{\alpha} + \frac{3}{4 \pi}\left({u}_{\rm th} - \frac{1}{4}
{u}_{\rm qm} \right) {\alpha}.
\end{eqnarray}
This system of equations admits two possible fixed points (see Fig.\
\ref{fig:flows}), viz.\ the trivial one at
${\nu}={r}={\alpha}=0$ which we already found above at $T=0$,
and a new one at ${\nu}={r}=0, {\alpha}={\alpha}^*$, with
\begin{equation} \label{fp}
{\alpha}^*=\frac{4\pi}{3 k_{\rm B} T}(1-\epsilon).
\end{equation}   
\noindent
Note that this fixed point exists even for a two-dimensional membrane
($\epsilon=0$).  The scaling field ${r}$ is found to be relevant, while
${\nu}$ is found to be irrelevant with respect to both fixed points.
Criticality is obtained by setting ${r}=0$.  The scaling field
${\alpha}$ behaves differently: it is irrelevant with respect to the
trivial fixed point, but relevant with respect to the new one.  The presence
of the unstable fixed point implies the existence of a crumpling transition
for a two-dimensional membrane at a critical temperature
\begin{equation}  \label{Tc}
T_{\rm c}= \frac{4\pi}{3 k_{\rm B}} \frac{1}{{\alpha}^*}.
\end{equation} 
For $T<T_{\rm c}$, $\alpha$ flows away from $\alpha^*$ and towards
the trivial fixed point $\alpha=0$. The correlation between the
normals to the surface is long-ranged, and the membrane remains flat.
For $T>T_{\rm c}$, on the other hand, ${\alpha}$ flows away from
${\alpha}^*$ in the other direction, that is ${\alpha} \to \infty$.
In this case, thermal fluctuations dominate.  The correlation between
the normals to the surface is short-ranged, and the membrane is found
to be crumpled.  As the temperature tends to infinity, the time
dimension shrinks to a point, making the integration $\int {\rm d}
\tau$ disappear from the action (\ref{action}).  This implies that the
parameter $\epsilon$ in the flow equations
(\ref{betanu})--(\ref{betaalpha}) must be set equal to
$\epsilon=\epsilon' + 1$ \cite{schmeltzer,stephens}, with
$\epsilon'=0$ for $D=2$.  We see that the fixed point (\ref{fp})
reduces in the limit $T \to \infty$ to the trivial one for a
two-dimensional membrane.  A similar nontrivial fixed point to the one
in (\ref{fp}) was found in Refs.\ \cite{peliti,leibler} for a
$(2+|\epsilon'|)$-dimensional membrane ($\epsilon' <0$) described by
the classical Canham-Helfrich model embedded in $3+|\epsilon'|$
dimensions. In our case, due to the presence of the extra time
dimension, the nontrivial fixed point (\ref{fp}) exists also in two
dimensions.

\section{Conclusions}

We have analyzed the large-scale behavior of a membrane subject to
thermal and quantum fluctuations.  At the absolute zero of
temperature, the membrane is found to be flat, and the vectors normal
to the surface are found to be strongly correlated.  This behavior,
being qualitatively different from that at high temperatures, led us
to investigate how these two regimes are related, by studying the
system at finite temperature. We found that it
exhibits a crumpling transition: above a critical temperature $T_{\rm
c}$, thermal fluctuations turn the membrane into a crumpled surface,
while for temperatures below $T_{\rm c}$, quantum fluctuations render
the membrane flat.

\acknowledgements
We thank M. C. Diamantini, V. Schulte-Frohlinde, and C. A. Trugenberger
for useful discussions.


\begin{thebibliography}{99}
\bibitem{qm1} S. V. Pereverzev, A. Loshak, S. Backhaus, J. C. Davis,
and R. E. Packard, Nature {\bf 388}, 449 (1997).
\bibitem{qm2} R. W. Simmonds, A. Loshak, A. Marchenkov, S. Backhaus,
S. Pereversev, S. Vitale, J. C. Davis, and R. E. Packard,
Phys. Rev. Lett. {\bf 81}, 1247 (1998).
\bibitem{Jerusalem} {\it Statistical Mechanics of Membranes and
Surfaces}, Proceedings of the Fifth Jerusalem Winter School for Theoretical
Physics, 1987/1988, edited by D. Nelson, T. Piran, and S. Weinberg
(World Scientific, Singapore, 1989).
\bibitem{helf3} W. Helfrich, J. Phys. (Paris) {\bf 46}, 1263 (1985).
\bibitem{peliti} L. Peliti and S. Leibler, Phys. Rev. Lett. {\bf 54},
1690 (1985). 
\bibitem{klein1} H. Kleinert, Phys. Lett. A {\bf 114}, 263 (1986).
\bibitem{forster} D. F\"{o}rster, Phys. Lett. A {\bf 114}, 115 (1986).
\bibitem{helf2} W. Helfrich, J. Phys. (Paris) {\bf 48}, 285 (1987).
\bibitem{forster2} D. F\"{o}rster, Europhys. Lett. {\bf 4}, 65 (1987).
\bibitem{fluid} H. Kleinert, J. Stat. Phys. {\bf 56}, 227 (1989).
\bibitem{cai} W. Cai, T. C. Lubensky, P. Nelson, and T. Powers,
J. Phys. II France {\bf 4}, 931 (1994).
\bibitem{DGT} P. G. de Gennes and C. Taupin, J. Phys. Chem. {\bf 86},
2294 (1982).
\bibitem{MW} N. D. Mermin and H. Wagner, Phys. Rev. Lett. {\bf 17}, 1133
(1969).
\bibitem{canham} P. B. Canham, J. Theor. Biol. {\bf 26}, 61 (1970).
\bibitem{helf1} W. Helfrich, Z. Naturforsch. {\bf 28c}, 693 (1973).
\bibitem{caroline} C. M. Fraser, Z. Phys. C {\bf 28}, 101 (1985).
\bibitem{memb} M. E. S. Borelli, H. Kleinert, and A. M. J. Schakel,
Phys. Lett. A {\bf 253}, 239 (1999).
\bibitem{Wilson} K. G. Wilson, Rev. Mod. Phys. {\bf 47}, 773 (1975).
\bibitem{Rivers} R. J. Rivers, {\it Path Integrals in Quantum Field
Theory} (Cambridge University Press, Cambridge, 1987).
\bibitem{Kapusta} J. I. Kapusta, {\it Finite-Temperature Field Theory}
(Cambridge University Press, Cambridge, 1989).
\bibitem{gradstein} I. Gradstheyn and I. M. Ryzhik, {\it Table of
Integrals, Series and Products} (Academic Press, Boston, 1980).
\bibitem{leibler} S. Leibler, in Ref.\ \cite{Jerusalem}.
\bibitem{schmeltzer} D. Schmeltzer, Phys. Rev. B {\bf 32}, 7512
(1985).
\bibitem{stephens} D. O'Connor and C. R. Stephens, Nucl. Phys. B {\bf
360}, 297 (1991).
\end{thebibliography}
\end{document}